\documentstyle[12pt]{article}
\begin{document}
\newcounter{numeris}
\begin{center}
{\large
ON THE QUANTUM EVOLUTION OF CHAOTIC SYSTEMS AFFECTED BY REPEATED FREQUENT
MEASUREMENT}\\[\normalbaselineskip]
B. Kaulakys\\[\normalbaselineskip]
Institute of Theoretical Physics and Astronomy,\\
A. Go\v stauto 12, 2600 Vilnius, Lithuania\footnote{Electronic mail
address: kaulakys@itpa.fi.lt}\\[2\normalbaselineskip]
\end{center}

\begin{center}
\parbox{6in}{{\bf Abstract.} We investigate the effect of repeated measurement
for quantum dynamics of the suppressed systems which classical counterparts
exhibit chaos. The essential feature of such systems is the quantum
localization phenomena strongly limiting motion in the energy space. Repeated
frequent measurement of suppressed systems results to the delocalization.
Time evolution of the observed chaotic systems becomes close to the
classical frequently broken diffusion-like process described by rate equations
for the probabilities rather than for amplitudes.}
\end{center}
\vskip 2\normalbaselineskip

\centerline{\large 1. Introduction}
\vskip \normalbaselineskip

Evolution of quantum systems, when they are not being observed, are described
by the Schr\" odinger equation. Measurements change abruptly the state of the
system and project it to the eigenstate of the measured observation. Repeated
frequent measurement can inhibit or even prevent the quantum dynamics.  This
is the essence of the quantum Zeno effect or the quantum watched pot [1-3].

The quantum Zeno effect means that time evolution of the system from an
eigenstate of the measured observable into a superposition of eigenstates is
inhibited by the measurement. It has been shown in a
recent experiment [3], that a variation of the quantum Zeno effect in a three
level system, originally proposed by Cook [2], can be realized. Fierichs and
Shenzle [4] have obtained the outcome of the experiment [3] on the basis of
the standard three-level Bloch equations without the {\it ad hoc} collapse of
the wave function.

Aharonov and Vardi [5] showed that frequent measurement can not only stop
the dynamics of a system, but also may induce a time evolution of a system: a
dense sequence of measurements along a presumed path induces a dynamics of the
system close to this arbitrary chosen trajectory (see, also [6]).

One of the specific features of the systems mentioned above is that they
consist only of the few (two or three) quantum states and are purely quantum.
The purpose of this report is to consider the influence of the repeated
frequent measurement on the evolution of the multilevel quasiclassical systems
which classical counterparts exhibit chaos. It has been established that the
chaotic dynamics of such, e. g. strongly driven by a periodic external field
non-linear
systems, is suppressed  of the quantum interference effect and results to the
quantum localization of the classical dynamics in the energy space of system
[7,8]. Thus, the quantum localization phenomena strongly limits the quantum
motion.

Here we investigate the effect of the repeated measurement for quantum
dynamics of suppressed systems. Our investigation is based on the mapping
equations of motion for systems with one degree of freedom perturbed by a
periodic force.
\vskip 2\normalbaselineskip

\centerline{\large 2. Quantum maps}
\vskip\normalbaselineskip

{}From the standpoint
of an understanding of manifestation of the classical chaos in the quantum
description of systems, the region of large quantum numbers is of greatest
interest. Here we can use the quasiclassical approximation and convenient
variables are the angle $\vartheta$ and the action $I$. Transition from the
classical to the quantum (quasiclassical) description can be undertaken
replacing $I$ by the operator $\hat I=-i{\partial\over\partial\vartheta}$
[9,10].
(We use the system of units with $\hbar=m=e=1$). One of the simplest systems
in which dynamic chaos and quantum localization of states can be obtained is a
system with one degree of freedom described by Hamiltonian $H_0(I)$
and perturbed by periodic kicks:
$$
H(I,\vartheta,t)=H_0(I)+K\cos\vartheta\sum\limits_j \delta(t-jT).\eqno(1)
$$
Here $T$ and $K$ are the period and parameter of the perturbation. The
intrinsic frequency of the unperturbed system is $\Omega=dH_0/dI$. In
particular for a linear oscillator we have $H_0=\Omega I$. For $H_0=I^2/2$ we
have the widely investigated rotator [7,8].

Integration of the Hamilton equations of motion over one perturbation period
leads to the classical map for the action and angle [9-11]:
$$
\left\{\begin{array}{c}
I_{j+1}=I_j+K\sin\vartheta_j\\
\vartheta_{j+1}=\vartheta_j+\Omega(I_{j+1})T.
\end{array}\right.\eqno(2)
$$

For derivation of quantum equations of motion we expand the state
function $\psi(\vartheta,t)$ of the system in quasiclassical eigenfunctions,
$\varphi_n(\vartheta)=e^{in\vartheta}/\sqrt{2\pi}$, of the Hamiltonian $H_0$,
$$
\psi(\vartheta,t)=(2\pi)^{-1/2}\sum\limits_na_n(t)e^{in\vartheta}.\eqno(3)
$$
Integrating the Schr\" odinger equation over a period $T$, we obtain the
following maps for the amplitudes [10]
$$
a_n(t_{j+1})=e^{-iH_0(n)T}\sum\limits_ma_m(t_j)J_{m-n}(K),~~~t_j=jT\eqno(4)
$$
where $J_m(K)$ is the Bessel function.

The form (4) of the map for the quantum dynamics is rather common: similar
maps may be derived, e. g., for the atom in a microwave field [10,12].
\vskip 2\normalbaselineskip
\centerline{\large 3. Dynamics}
\vskip \normalbaselineskip

Classical dynamics of the
system described by map (2) in the case of global distinct stochasticity is
diffusion-like with the diffusion coefficient in the $I$ space resulting from
(2),
$$
B(I)=K^2/4T.\eqno(5)
$$
{}From equations (4) we have the transitions probabilities $P_{n,m}$ between
$n$
and $m$ states during the period $T$:
$$
P_{n,m}=J^2_{m-n}(K).\eqno(6)
$$
Using the expression $\sum_nn^2J^2_n(K)=K^2/2$ and approximation of
uncorrelated transitions we may formally evaluate the local quantum diffusion
coefficient in the $n$ space (see also [13,14]):
$$
B(n)={1\over 2T}\sum\limits_m(m-n)^2J_{m-n}^2(K)={K^2\over 4T}.\eqno(7)
$$
Therefore, the local quantum diffusion coefficient coincides with the classical
one (5).

However, it turns out that such a quantum diffusion takes place only for some
finite time $t^*$ after which an essential decrease of the diffusion rate is
observed. Such a behaviour of the quantum systems in the region of strong
classical chaos was called "the quantum suppression of classical chaos". This
phenomenon turns out to be typical for models (1) with the nonlinear
Hamiltonians $H_0(I)$ and other quantum systems. Thus, the diffusion
coefficient (7) derived in the approximation of uncorrelated transitions (6)
does not describe the true quantum diffusion in the energy space.
The quantum interference effect is essential and results to the quantitatively
different dynamics. This is due to the pseudorandom nature of the phases
$H_0(n)T$ in equation (4) as functions of $n$ (but not $j$). Replacing
$\exp[-iH_0(n)T]$
in (4) by $\exp[-i2\pi g(n)]$, where $g(n)$ is a sequence of random numbers
that are uniformly distributed in the interval [0,1] we observe the quantum
localization as well [10]. However, the essential point is the independence
of the phases $H_0(n)T$ on the step of iteration or time $t_j$. For the random
phases as functions of iteration steps we observe the unlimited motion in the
$n$ space.
\vskip 2\normalbaselineskip
\centerline{\large 4. Influence of measurements}
\vskip \normalbaselineskip
Each measurement of the energy projects the system into one of the energy
eigenstates with definite $n$. Therefore, if we make a measurement before the
next kick we will find the system in the states $\varphi_m$ with the
appropriate probabilities $P_m=|a_m|^2$. After the measurement the phase of
the amplitude $a_m(t_j)$ is random. Thus, if there are measurements of the
system's state before or after each kick, we have from equations (4) the
system of rate equations for probabilities
$$
P_n(t_{j+1})=\sum\limits_mJ^2_{m-n}(K)P_m(t_j).\eqno(8)
$$

The rate equations (8) result to the unlimited diffusion-like motion in the
$n$ space with the diffusion coefficient (7). Therefore, the frequent
measurement of the quantum system described by map (4), which exhibits the
quantum suppression of classical chaos, results to the diffusion-like motion,
similar to the classical dynamics.

The analogous conclusion may be drawn and for other systems, as well.
\newpage
\centerline{\large 5. Conclusions}
\vskip \normalbaselineskip
Repeated frequent measurement of the simple two- or three-state system inhibit
or even prevent the quantum dynamics of the system. The similar measurement of
the multilevel quasiclassical system with quantum suppression of classical
chaos results to the dynamics described by rate equations for probabilities
rather than for amplitudes and causes the delocalization of the states
superposition. The quantum dynamics of such chaotic systems affected by
repeated frequent measurement resembles the classical diffusion-like motion.

It should be noted that the same effect may be derived without the {\it ad hoc}
collapse hypothesis but from the quantum theory of irreversible processes.
Even the simplest detector follows irreversible dynamics due to the coupling to
the multitude of vacuum modes. The quantum system attains irreversibility due
to
the interaction with the detector and the phases of the amplitudes or the
off-diagonal matrix elements of the density matrix decay steadily (see for
analogy [4]). Measurement of such a type may also result to the disappearance
of the effect of the quantum suppression.
Therefore, the quantum evolution of
chaotic system interacting with the environment or detector is more
classical-like than the evolution of the idealized isolated system.
\vskip 2\normalbaselineskip
\centerline{\large Acknowledgement}
\vskip \normalbaselineskip

The research described in this publication was made possible in part by Grant
No. LAA000 from the International Science Foundation.
\vskip 2\normalbaselineskip
\centerline{\large References}
\vskip \normalbaselineskip
\begin{list}{\arabic{numeris}.\hfil}%
{\usecounter{numeris}\labelwidth=0.20in\leftmargin=0.25in\labelsep=0.05in%
\itemsep=0in\parsep=0in}
\item B. Misra and E. C. G. Sudarshan, J. Math. Phys. {\bf 18}, 756 (1977).

\item R. J. Cook, Phys. Sripta {\bf T21}, 49 (1988).

\item W. M. Itano, D. J. Heinzen, J. J. Bollinger and D. J. Wineland,
Phys. Rev. A {\bf 41}, 2295 (1990).

\item V. Frerichs and A. Schenzle, Phys. Rev. A {\bf 44}, 1962 (1991).

\item Y. Aharonov and M. Vardi, Phys. Rev. D {\bf 21}, 2235 (1980).

\item T. P. Altenm\" uller and A. Schenzle, Phys. Rev. A {\bf 48}, 70 (1993).

\item G. Casati, B. V. Chirikov, D. L. Shepelyansky and I. Guarneri,
Phys. Rep. {\bf 154}, 77 (1987).

\item F. M. Izrailev, Phys. Rep. {\bf 196}, 299 (1990).

\item G. M. Zaslavskii, Stochastic Behaviour of Dynamical Systems, Moscow:
Nauka (1984).

\item V. G. Gontis and B. P. Kaulakys, Liet. Fiz. Rink. {\bf 28}, 671 (1988)
[Sov. Phys.-Collec. {\bf 28}(6) 1 (1988)].

\item A. J. Lichtenberg and M. A. Lieberman, Regular and Stochastic Motion,
New York: Springer (1983).

\item R. V. Jensen, S. M. Susskind and M. M. Sanders, Phys. Rep. {\bf 201},
1 (1991).

\item V. Gontis and B. Kaulakys, J. Phys. B {\bf 20}, 5051 (1987).

\item B. Kaulakys, V. Gontis, G. Hermann and A. Scharmann, Phys. Lett.
{\bf A159}, 261 (1991).
\end{list}
\end{document}